# Gamma Ray Cherenkov-Transition Radiation

M.A. Aginian[1], K.A. Ispirian[1], M. Ispiryan[2]

[1] Alikhanian National Laboratory (Yerevan Physics Institute), Br. Alikhanian 2, Yerevan, 0036, Armenia

[2] University of Pennsylvania, Philadelphia, Pennsylvania, 19104, USA

**Abstract**

The production of gamma-ray Cherenkov-transition radiation (GCTR) by charged particles in the photon energy region 0.8÷2 MeV is studied theoretically using the results of the recent discovery that in the above mentioned region the dielectric constant or the refraction index of some materials is greater than 1 due to Delbruck scattering on Coulomb field of nuclei. Using the results of the carried out numerical calculations, the possibility of observing GCTR and some of its applications are discussed.

## 1. Introduction

It is well known that when a charged particles passes through a radiator medium with refraction index $n(\omega)>1$, it produces Cherenkov radiation (CR) of optical and softer photons under certain angle $\theta = \arccos(1/\beta n)$ if the particle's velocity is greater than a certain threshold value, $V > V_{thr} = c/n$. P. Cherenkov, I.M. Frank and I. Tamm received a Nobel Prize for the discovery [1] and development of the theory [2] of CR.

It is much less is known that theoretically [3] and experimentally [4-7] (see also the reviews [8, 9]) it was shown that in some materials X-ray Cherenkov radiation (XCR) is produced in narrow soft X-ray regions close to K-, L- and M-edges at photon energies at $\hbar\omega<1$ keV where $n(\omega)>1$. Though in [10-12] and some other works one can find hints on the existence of the CR in the region from soft X-rays up to gamma-ray regions, only in [3] it has been calculated and shown for the first time that for some materials the dielectric constant can be greater than 1. The theory of XCR was been developed by using complex values of $\varepsilon(\omega)$, or $n(\omega)$: namely, $\varepsilon(\omega) = \varepsilon'(\omega) + i\varepsilon''(\omega) = 1 + \chi'(\omega) + i\chi''(\omega)$ or $n(\omega) = \sqrt{\varepsilon(\omega)} = 1 + \delta(\omega) + i\Delta(\omega)$ in the CR or X-ray transition radiation (XTR) theories.

In 2012 study of the scattering of 0.5÷2 MeV gamma quanta in the work [13] it has been shown that the refraction index of silicon becomes slightly larger than 1 ( $\delta = \chi'/2 \approx 10^{-10} \div 10^{-9} > 0$) in the region 0.75÷1.96 MeV. The authors of [13] explain this observation with the help of QED nonlinear process of Delbruck scattering, which takes place as photon-photon scattering by production of virtual electron-positron pairs. This fact can find wide application in gamma ray optics for focusing of the gamma ray beams [13].

The purpose of the present short work is: using the results of [13], make prediction and study the gamma ray Cherenkov radiation. Since the main results are obtained using XTR theory and for some conditions one may not consider CG and TR independently each from other [8, 10], it is reasonable to name this radiation gamma-ray Cherenkov-transition radiation (GCTR).

**2. Theory of GCTR Produced in a Plate on the Basis of XTR theory**

Now, following [3-9], we will consider GCTR using the Ginzburg-Frank-Garibian theory on XTR produced in a plate. The validity of this theory is better justified than that of Frank-Tamm theory of CR, because in reality one has a plate radiator in vacuum or air, and the GCTR photons are detected at some distance after the radiator. Instead of the forward TR formula in the general form as in [11] using the formula (2.21) of [14], or coinciding formula (3.27) of [8] for the spectral-angular distribution of XTR produced by an ultra-relativistic charged particle in a plate, one can easily derive the spectral-angular distribution of the number of GCTR photons by inserting the values of the complex $\chi$:

$$\frac{d^2N}{d\hbar\omega d\theta} = \frac{2\alpha\theta^3}{\pi\hbar\omega} \frac{(\chi'^2 + \chi''^2)}{(\theta^2 + \gamma^{-2})[(\theta^2 + \gamma^{-2} - \chi')^2 + \chi''^2]} \times \left\{ \left[1 - \exp\left(-\frac{L_{rad}}{2L_{abs}}\right)\right]^2 + 4\exp\left(-\frac{L_{rad}}{2L_{abs}}\right)\sin^2\frac{\hbar\omega L_{rad}(\theta^2 + \gamma^{-2} - \chi')}{4\hbar c} \right\}, \quad (1)$$

where $\alpha = e^2/\hbar c$ is the fine structure constant, $\gamma = E/mc^2$ is the relativistic factor of electrons, $L_{rad}$ is the thickness of the radiator and $L_{abs} = 1/\mu = \hbar c/\chi''\hbar\omega$ ($\mu$ is the absorption linear coefficient) is the photon absorption length.

Postponing the discussion of the particular cases of the more or less general formula (1) when $L_{rad} \gg L_{abs}$ and $L_{rad} \ll L_{abs}$, as well as of discussion of the notion of formation length of GCTR, $L_{form}^{GCTR}$, for a later publication, in this short letter we shall discuss the characteristics of GCTR with the help of numerical calculations using (1) in order to prepare an experimental study of GCTR.

Let us remind that our purpose is to integrate (1) over a) $\hbar\omega$ to obtain $dN_{GCTR}/d\theta$ and b) over $\theta$ to obtain $dN_{GCTR}/d(\hbar\omega)$ and finally c) over $\hbar\omega$ and $\theta$ in order to obtain the total number $N_{GCTR}$ of GCTR photons. One cannot integrate (1) analytically over $\hbar\omega$. In principle, one can integrate (1) over $\theta$ analytically, however, the obtained result has very complicated form with integral sine and cosine functions. Therefore, in further calculations we will carry out only numerical integrations.

## 3. Numerical Results and Discussion

To begin, by integrating (1) over $\hbar\omega$ and $\theta$, we have calculated the dependence of the total number of photons $N_{GCTR}$, produced by electrons of various energies in Si plate, upon the plate thickness $L_{rad}$. As in the case of XTR at small thicknesses, $N_{GCTR}$ increases proportionally to the square of $L_{rad}$ with the increase of $L_{rad}$, then this dependence weakens, yielding a maximum at thickness of a few cm. Therefore, it is reasonable to consider all the further Si radiators having thickness $L_{rad}=1$ cm, i.e., less than the radiation length of Si equal to $X_0^{Si}=9.36$ cm when the primary electron energy degradation in the radiator is small. We will consider the electron energy to be $E=20$ GeV - an energy available at SLAC and CERN SPS and which is, as it will be shown below, slightly higher than the GCTR threshold energy providing enough number of produced GCTR photons. In this work the calculations are performed for photon energy interval 800÷2000 MeV, interpolating the more confident, but scarce experimental data on $\chi'=\delta/2$ for three energies and assuming that $\chi'=0$ at ~2.5 MeV.

Fig. 1 shows the angular distribution of GCTR photon number after the integration of (1) in the photon energy interval 800÷2000 MeV for various θ angles. As it is seen, the angular distribution is broader than that of CR and XCR [8]. Similar distributions, calculated at various electron energies, show that with the increase of E the peak of angular distribution occurs at smaller angles.

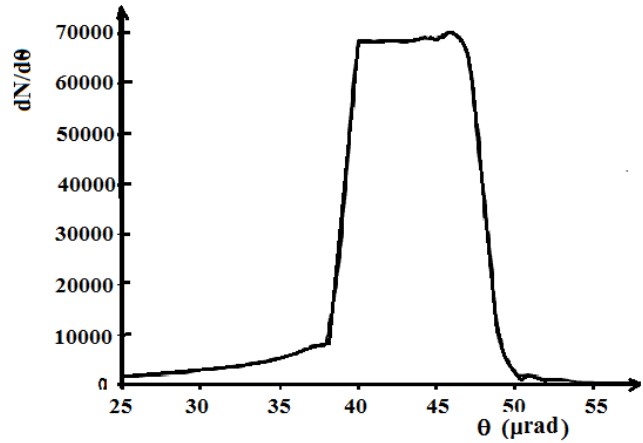

Fig. 1. The angular distribution of GCTR.

The spectral distribution of the number of GCTR photons obtained, as in the case of XTR, after integration of (1) over θ in the interval from 0 up to ∞ (because (1) sharply decreases with the increase of θ) is shown in Fig. 2 (solid curve).

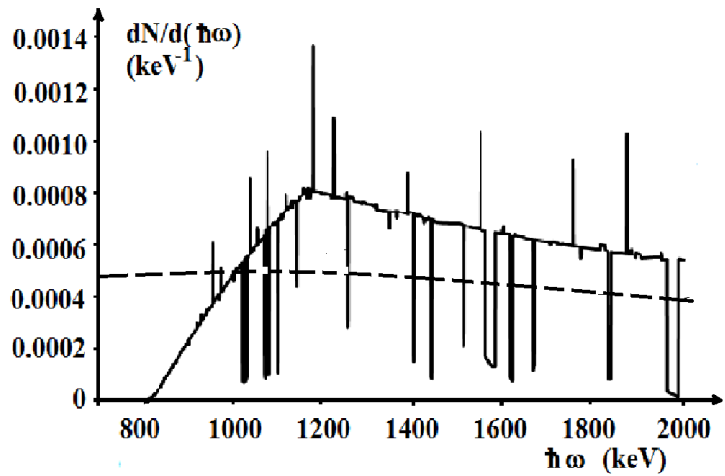

Fig. 2. The spectral distributions of GCTR (solid curve, 1) and bremsstrahlung (dashed curve, 2).

At the entrance and exit interfaces of the radiator or due to sine in formula (1) there are splashes and drops in the spectral distribution at certain values of photon energies. The dotted curve 2 in Fig. 2 shows the ten-fold increase in the magnitude of the spectral distribution of bremsstrahlung produced in 1 cm thick Si plate taking into account the longitudinal density or Ter-Mikaelian effect (see [15]). Here it is worthy to note a) that the density effect is manifested at photon energies less than $\omega_P\gamma \approx 1200$ $keV$ and it is not necessary to take into account the Landau-Pomeranchuk-Migdal (LPM) effect, since LPM effect is manifested at much higher electron energies; b) as it is has been shown for the first time in [16], the number of the produced (without absorption) resonance transition radiation (RTR) photons in transition radiation detectors (TRD) from a stratified radiator exceeds the background bremsstrahlung only in a relatively narrow region $\Delta(\hbar\omega) \approx (2\div 20)$ keV, and this allows to study XTR with the help of electrons and construct TRD detectors which found wide application for particle identification. As it follows from Fig. 2, the GCTR intensity exceeds the background bremsstrahlung background in a much wider interval $\Delta(\hbar\omega) \approx (800\div 2000)$ keV; c) the total number of GCTR photons at $E=20$ GeV, obtained by numerical integration of the curve in Fig. 1 over $\theta$ or of the solid curve 1 in Fig. 2 over $\hbar\omega$, is equal to $N_{GCTR}=0.7$, while the total number of the background bremsstrahlung photons obtained by numerical integration of Ter-Mikaelian's formula or dotted curve of Fig. 2 is equal to $N_{Br}=0.06$.

The dependence of the number of GCTR photons produced in the case of arrangement with same parameters as in Figs. 1 and 2 upon electron energy $E$, calculated by the numerical integration of (1) in the photon energy interval $\hbar\omega=(800\div 2000)$ keV and all possible angular values of $\theta$, is shown in Fig. 3. Let us note that usually the energy dependence of RTR photons of the TRDs [17] increases sharply beginning from $\gamma=10^2$ and saturates beginning from $\gamma=10^3$. As it is seen from Fig. 3, the behavior of GCTR compared with RTR is shifted to higher energies approximately by 1 order of magnitude, i.e., "the threshold energy of GCTR is higher". Besides, the number of produced RTR photons (without taking into account the absorption) at very high "saturation" energies is approximately ~ 11 (see [16]), while the "saturation" number of GCTR photons, as it follows from Fig. 3, is ~1. One can understand this fact by the above discussed broader interval of GCTR compared with the narrower interval of RTR.

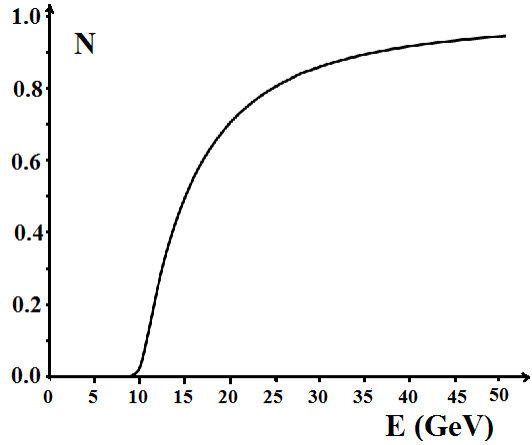

Fig. 3. The energy dependence of the number of the GCTR photons (see the text).

As it follows from the above results, one can detect the GCTR photons, as predicted in this work, by using electrons with energies higher than ~10 GeV when a few GCTR photons are produced with the help of a detector schematically shown in Fig. 4. This arrangement is similar to the one used for the study of LPM and Ter-Mikaelian effects at SLAC [18,19].

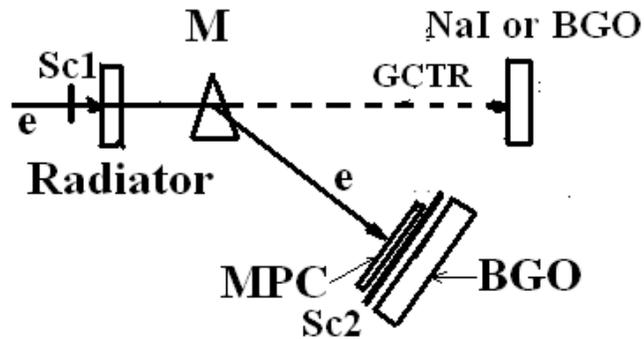

Fig.4. A possible experimental arrangement.

The primary and secondary electrons are detected with the help of thin scintillation counters Sc1 and Sc2. They pass through a ~0.1$X_0$=1 cm thick Si radiator, and after deflection by the magnet M, are detected by multiwire proportional chamber (MPC) and thick BGO calorimeter which measures also the electron energy after radiation. Almost all the GCTR and bremsstrahlung photons produced per one electron in Si are detected with the help of the forward NaI(Te) or BGO spectrometer measuring the spectral distribution of GCTR.

In all the above calculations polynomial approximations of $\chi', \chi''$ were used. The used interpolations allow us to estimate the accuracy of the calculations to be not better than 20%.

We now come to conclusions. In this work, a new type of radiation, GCTR, it is predicted and studied. The radiation can find several applications. a) Taking into account the difficulties d experimenting in gamma-ray region (see [12]), the GCTR may serve as an easier method for finding materials with $n>1$. b) Since the GCTR threshold energy is high, $\gamma_{thr} >> 10^4$, the GCTR can serve for particle identification. Indeed, as it is well known, the optical CR, the threshold of which is given as $\beta_{thr} = 1/n$, has a low value of $\gamma_{thr}$. TRDs have a "threshold" at $\gamma_{thr} \approx 10^2 \div 10^3$. Only by detecting the Compton-scattered XTR photons it is possible to increase the $\gamma_{thr}$ [17] slightly. Now, as it has been shown above, the GCTR has $\gamma_{thr} \approx 2 \times 10^4$. Therefore, just as in case of XTR, one can use the GCTR for particle identification at higher energies. c) Since the number of the GCTR photons produced per electron is of the order of 1, one can use the mechanism of GCTR to produce intense beams of gamma quanta beams with energies 1÷2 MeV. And, finally, d) One can use such intense gamma beams for gamma ray optics and nuclear physics, in particular, for study of giant resonances.

The authors would like to thank the physicists who have reviewed the article; their opinion helped to correct and improve it.